\documentclass[prl,showpacs,twocolumn,nobibnotes,nofootinbib,floatfix]{revtex4}

\usepackage{amsmath}
\usepackage{amssymb}
\usepackage{graphicx}
\usepackage{amsthm}

\DeclareMathOperator{\Tr}{Tr}
\DeclareMathOperator{\Ker}{Ker}
\DeclareMathOperator{\diag}{diag}

\newcommand{\R}{\ensuremath{\mathbb R}}

\theoremstyle{definition}
\newtheorem{theorem}{{Theorem}}

\begin{document}

\title{Concurrence of Stochastic 1-Qubit Maps}

\date{\today}

\author{Meik Hellmund}
\affiliation{Mathematisches Institut, Universit{\"a}t Leipzig,
Johannisgasse 26, D-04103 Leipzig, Germany}
\email{Meik.Hellmund@math.uni-leipzig.de}

\author{Armin Uhlmann}
\affiliation{Institut f{\"u}r Theoretische Physik, Universit{\"a}t Leipzig,
Vor dem Hospitaltore 1, D-04103 Leipzig, Germany}
\email{Armin.Uhlmann@itp.uni-leipzig.de}

\begin{abstract}
  Explicit expressions for the concurrence of all positive
  and  trace-preserving (``stochastic'') 1-qubit  maps are
  presented. By a new method we find  the relevant convex
  roof pattern. We conclude   that two component optimal
  decompositions always exist.

  Our results can be transferred to $2 \times n$-quantum
  systems providing the concurrence for all
  rank two density operators as well as a  lower bound
  for their entanglement of formation.

\end{abstract}


\maketitle

\section{Introduction}
In quantum physics a  system in a pure state $\pi=|\psi\rangle\langle\psi|$ 
may have subsystems 
in states which are not pure but mixed. These mixed substates
are  typically correlated in a  non-local and non-classical way.
The use of this  phenomenon of entanglement
as a resource for communication 
and computation is a main feature  of
quantum information theory \cite{NC00}. This makes the 
search for a quantitative understanding and 
characterization of entanglement a central 
issue \cite{horodecki-2007,bengtsson06}.
Entanglement measures  ought to describe
single-use or asymptotic capabilities of quantum systems and channels
just as the von Neumann entropy $S(\rho)=- \Tr \rho\log\rho$ is an
asymptotic measure for information content. 
  They are, similar to  entropy, 
non-linear and unitarily invariant functions on the space of states.

Bennett et al. \cite{BenFucSmo96} introduced the {\it
  entanglement of formation} $E_F(\rho)$ as the asymptotic number of 
ebits (maximally entangled qubit pairs)  needed to prepare the entangled 
bipartite state $\rho$ 
by local operations and classical communication (LOCC) and showed that
\begin{equation}
  \label{eq:13}
  E_F(\rho) = \min 
\sum p_j\; S\left(\Tr_B (\pi_j )\right)
\end{equation}
where $\Tr_B$ is the partial trace over one of the two subsystems 
and the minimium is taken  over all possible 
convex ($\sum p_j=1, p_j>0$) decompositions of the state $\rho$ 
into pure states 
\begin{equation} \label{t2.2}
\rho = 
\sum p_j\, \pi_j, \quad \pi_j \, \hbox{ pure.}
\end{equation}
Closed formulas for the entanglement of formation, i.e., 
analytic solutions to the global optimization problem (\ref{eq:13}) 
are only known for certain classes of highly symmetric states 
\cite{TV00,vollbrecht-2000} and for the  case of a pair of qubits
($2\times 2$ system). In the latter case, the analytic formula 
for the entanglement of formation
was obtained first for special states \cite{BenFucSmo96,HilWoo97} and later
proved for all states of a qubit pair \cite{Woo97}.  
It expresses $E_F(\rho)$ in terms of another entanglement measure
$C(\rho)$ which was named {\it concurrence} in \cite{HilWoo97}.
The concurrence appeared to be an interesting quantity in itself \cite{Woo01}.
Many authors, e.g.  \cite{lozinski-2003,chen05,Fei07}, 
have obtained bounds for the 
 concurrence of larger bipartite systems.  

In the present paper we obtain analytic expressions for the 
concurrence for general stochastic 1-qubit maps and therefore for
{\em general $2\times n$ bipartite systems 
 provided the input state $\rho$ has rank two.}
For this we employ the  
{\it convex roof} construction \cite{BNU96,Uh98} as a way to study 
global optimization problems of the type (\ref{eq:13}). Our main results are 
given by Theorems~2 and~3.

Let $\Phi$ be a positive and trace-preserving (i.~e., stochastic) map
from a general quantum system into a 1-qubit-system. This setup 
includes as
special case the partial trace $\Tr_B$ which maps states of a bipartite
$2\times n$  
system to states of the subsystem.
For pure input states $\pi=|\psi\rangle\langle\psi|$ the
concurrence is defined as
\begin{equation}
  \label{eq:1}
  C_\Phi(\pi) = 2\sqrt{\det \Phi(\pi)}
\end{equation}
and for a general mixed input state $\rho$ one defines
\begin{equation} \label{t2.1}
C_\Phi(\rho) = \min \sum p_j\, C_\Phi(\pi_j) \; ,
\end{equation}
where the minimum is again taken over all possible convex 
decompositions into pure states.
Let us consider the case where  $\rho$ has rank 2 and is therefore 
supported by a 2-dimensional input subspace. 
Then we have  to consider in (\ref{t2.2}) only pure states
supported in the same 2-dimensional supporting input space. By
unitary equivalence we are allowed to identify input and output
subspaces. Hence, calculating the concurrence  of a
rank two density operator $\rho=\sum_{i,j=1}^2 \rho_{ij}|v_i\rangle\langle
v_j|$ 
of a $2\times n$ system is 
equivalent to computing the
concurrence of a certain 1-qubit stochastic map. This map is 
completely positive and 
explicitely 
given by
$\Phi(\rho) = \sum_{i,j}\rho_{ij} D_{ij}$ with $D_{ij}= \Tr_B |v_i\rangle\langle
v_j|$.


However, our construction of the concurrence works for all stochastic
1-qubit maps, not only for completely positive ones. It is therefore
suggestive, but not the topic of the present paper, to ask for
applications to the entanglement witness problem \cite{horo96}.

In some cases the convex roof for the concurrence appears to be a {\it flat
  convex roof}. In these cases optimal decompositions for the concurrence 
also provide optimal decompositions for the entanglement of formation 
and therefore $E_F(\rho)$ can be
expressed as a  function of the  concurrence $C(\rho)$, exactly 
 as in the case of a pair of qubits \cite{Woo97}.  If the roof of the
 concurrence is not flat, our results for the concurrence provide 
 a lower bound for the entanglement of
 formation.

We illustrate our procedure by explicit formulas for the cases of
bistochastic and of  axial symmetric 1-qubit maps. In both cases the result
is of a surprising transparency.


\section{The convex roof construction}
In the following, all linear
combinations are understood as convex combinations, i.e., the $\{p_j\}$
always satisfy $\sum p_j=1$ and $p_j>0$.
Solutions to the optimization problem
eq.~(\ref{t2.1},\ref{t2.2})
can be characterized as so-called {\it convex roofs}: Let
$\Omega$ denote the convex set of density operators $\rho$
and let $g(\pi)$ be a continuous real-valued function on 
the set of pure states.
\begin{theorem}[see \cite{BNU96,Uh98}]
  \label{thmRoof}
There exists exactly one  function $G(\rho)$ on $\Omega$ which can be
characterized uniquely by each one of the  following two properties:
\begin{enumerate}
\item $G(\rho)$ is the solution of the optimization problem
  \begin{equation}
    \label{eq:5}
    G(\rho) = \min_{\rho=\sum p_j\,\pi_j}\sum p_j\, g(\pi_j).
  \end{equation}
\item $G(\rho)$ is convex \cite{Roc70} and a {\it roof},  
i.e., for every $\rho\in \Omega$
  exists an extremal  decomposition $\rho=\sum p_j \pi_j$ such that
  \begin{equation}
    \label{eq:6}
    G(\rho) = \sum p_j\, g(\pi_j) \; .
  \end{equation}
\end{enumerate}
Furthermore, given $\rho$, the function $G$ is  linear
 on the convex hull of all those
pure states $\pi_j$ which  appear in the decomposition (\ref{eq:6}) of
$\rho$. Therefore, $G$ provides a foliation of $\Omega$ into leaves such
that   a) each leaf is the convex hull of
some pure states and b) $G$ is  linear on each leaf.
If $G$ is not only linear but even constant on each leaf,  it is a
{\it flat roof}.
\end{theorem}

\section{Stochastic 1-qubit maps}
The space ${\cal M}_2$ of hermitian 2$\times$2 matrices
$
{ \rho}= \left(
\begin{smallmatrix}
  x_{00} & x_{01}\\ x^\ast_{01} & x_{11}
\end{smallmatrix}\right)
$
is isomorphic to  Minkowski space $\R^{1,3}$ via
   \begin{eqnarray}
     \label{eq:2} {\bf x} = (x_0,\vec x)\quad \Longleftrightarrow\quad
  { \rho}  &=&  \frac{1}{2} (x_0 I+\vec x\cdot \vec\sigma) \\
&=&\frac{1}{2}
  \begin{pmatrix}
    x_0+x_3 & x_1+i x_2\\ x_1-ix_2 & x_0-x_3
  \end{pmatrix}\nonumber.
\end{eqnarray}

We have  $ \det \rho = \frac{1}{4}(x_0^2 -x_1^2 -x_2^2-x_3^2
)=\frac{1}{4}{\bf x}\cdot{\bf x}$ where the dot between 4-vectors denotes the
Minkowski space inner product
 and
$    \Tr \rho  =  x_0$.   Therefore the cone of positive matrices
is just the forward light cone and
the state space $\Omega$ of a qubit, the Bloch ball,
is the intersection
of this cone with the hypersurface $V$ defined by   $x_0=1$.
In this picture 
mixed states correspond to time-like vectors
and pure states to light-like vectors, both normalized to $x_0=1$.

A trace-preserving positive linear map 
 $\Phi: {\cal M}_2 \rightarrow
{\cal M}_2$ can be parameterized as \cite{KR00}
\begin{equation}
  \label{eq:3} \Phi(\rho)=
  \Phi\left(\frac{1}{2}(x_0 I + \vec x \cdot\vec \sigma)\right) =
\frac{1}{2}\left(
x_0I + ( x_0 \vec t+{\bf \Lambda}\vec x) \cdot \vec \sigma\right)
\end{equation}
where $\bf \Lambda$ is a 3$\times$3 matrix and
$\vec t$ a 3-vector.

We consider the  quadratic form $q$  on ${\cal M}_2$ defined by
\begin{widetext}
\begin{equation}
  \label{eq:4}
  q^\Phi_{w}({\bf x}) =
4(\det\Phi(\rho)-w \det \rho) = \Phi({\bf x})\cdot \Phi({\bf x})
  -w\, {\bf x}\cdot {\bf x} = \sum_{i,j=0}^4 q_{ij} x_i x_j
\end{equation}
  \end{widetext}
where $w$ is some real parameter.
For pure states, i.e., on the boundary of the Bloch ball where ${\bf x}\cdot
{\bf x}=0$, the form $q({\bf x})$ equals the square of the concurrence
$C=2\sqrt{\det\Phi(\rho)}$.

Furthermore, we denote by $Q$  the linear map
 $Q: x_i\mapsto \sum q_{ij} x_j$ corresponding
to the quadratic form $q$ via polarization:
\begin{equation}
  \label{eq:15}
  Q^\Phi_w = Q^\Phi_0 - w\, \eta_{ij} = 
  \begin{pmatrix}
    1-|\vec t|^2-w & -\vec t {\bf\Lambda} \\
     -(\vec t {\bf\Lambda})^T  & 
w\, {\bf I}-{\bf\Lambda}^T{\bf\Lambda}
  \end{pmatrix} 
\end{equation}
The central result of this paper are the following two statements:
\begin{theorem}\label{thm1}
Let the  quadratic form $q$ and therefore the
matrix $Q$ be positive  semidefinite and  degenerate, i.e.,
$Q\ge 0$ and $\dim\Ker Q>0$. If $\Ker Q$ contains a non-zero
vector ${\bf n}$ which is
space-like or light-like, ${\bf n}\cdot{\bf n}\le0$, then
$q^{1/2}$ is a {\it convex roof}.
Furthermore, this  roof is {\it flat} if such an ${\bf n}$ exists with
$n_0=0$.
\end{theorem}
\begin{theorem}\label{thm2}
For every positive trace-preserving map $\Phi$
 exists a unique value $w_0$ for the parameter $w$
such that the conditions  of
Theorem~2 are fulfilled.  Therefore  the concurrence of an arbitrary
stochastic 1-qubit map $\Phi$ is given  by
 $C_\Phi(\rho)=\sqrt{q^\Phi_{w_0}(\rho)}$.

\end{theorem}
Let us sketch the proof of Theorems~\ref{thm1} and \ref{thm2}.
The square root $\sqrt{q}$ of a positive semidefinite form $q$
on a linear space
provides a seminorm on this space  and
is therefore convex. According to Theorem~\ref{thmRoof} we need to show
that it is also a roof, i.e., there is a foliation of the space
into leaves such that $q^{1/2}$ is linear on each leaf.
Let ${\bf n}=(n_0,\vec n)$ be a non-zero vector in $\Ker Q$.
Then for all vectors $\bf m$ we have
\begin{equation}
  \label{eq:7}
q({\bf m} +{\bf n}) = ({\bf m} +{\bf n})Q ({\bf m} +{\bf n})
= {\bf m} Q {\bf m} = q({\bf m}).
\end{equation}
  Let us start with the case where ${\bf n}$ can be chosen to have
  $n_0=0$.  Then $\vec n$ gives a direction in $V$ along which $q$
is constant. Therefore, $\sqrt{q}$ is a  {\it flat convex roof}.

\begin{figure}[h!t]
\label{fig:2}
\includegraphics[]{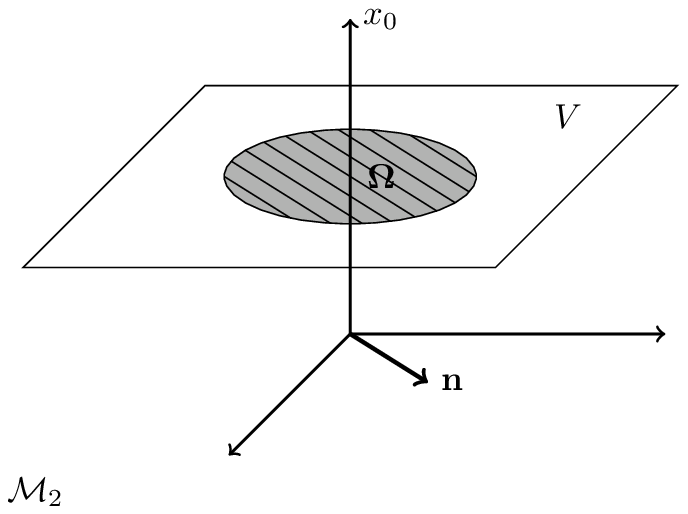}
\caption{The embedding of the Bloch ball into ${\cal M}_2$ and its
 foliation  by a flat convex roof.}
\end{figure}

Let us now consider the case where $\Ker Q$ does not contain a vector ${\bf
  n} $ with $n_0=0$. Then we have  $\dim\Ker Q=1$ and
this line intersects $V$ in
one point which we call ${\bf n}$. Every other point $\bf m$ in $V$
can be connected to the point $\bf n$ by a line lying in $V$.
Then
$q^{1/2}$ is linear along
the half-line $\R^+\ni s\mapsto s{\bf m} + (1-s) {\bf n}$ since
\begin{eqnarray}
  \label{eq:8}
  q\left(s{\bf m} + (1-s) {\bf n}\right) &=& (s{\bf m} + (1-s) {\bf n})Q
(s{\bf m} + (1-s) {\bf n}) \nonumber \\ & =&               s^2 q({\bf m})
\end{eqnarray}
This concludes the proof of Theorem~\ref{thm1}.
\begin{figure}[h!t]
\label{fig:3}
\includegraphics[]{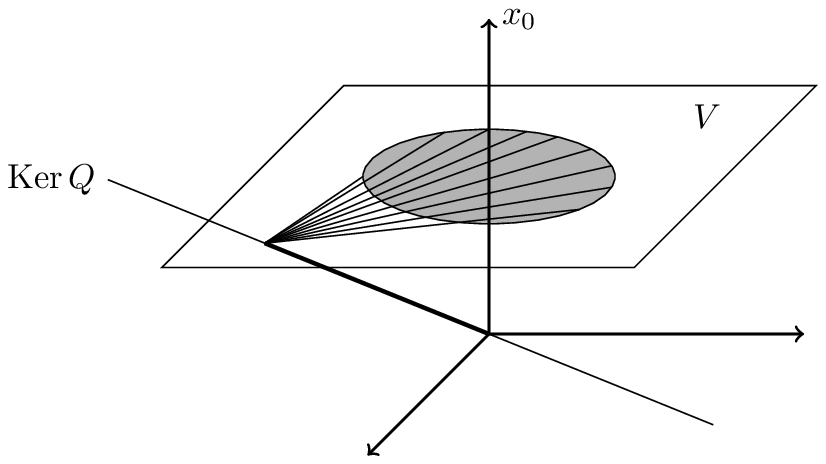}
\caption{The foliation of the Bloch ball in the case $n_0\ne 0$.}
\end{figure}

For the proof of Theorem~\ref{thm2} we note that
the space ${\cal P}$ of stochastic maps is itself a convex space. It can be
parameterized as follows
\cite{sudarshan76}:
Let $\vec \xi$ be a unit 3-vector and
$\alpha, \beta, \omega_1,\omega_2,\omega_3$ be
parameters taking values between zero and one:
$0\le\alpha\le1;\; 0\le\beta\le1;\; 0\le \omega_1\le \omega_2\le \omega_3=1$.
 With the abbreviation $\nu=\sqrt{\sum_{i=1}^3\xi_i^2\, \omega_i^2}$
we can represent stochastic maps (\ref{eq:3}) 
  up to
orthogonal transformations by $\vec t = (t_1,t_2,t_3)$, 
${\bf \Lambda}=\diag(\lambda_1,\lambda_2,\lambda_3)$ where 
 \begin{eqnarray}
   \label{eq:1x}
   t_i &=& \beta\, \xi_i\, (1-\alpha \omega_i^2)\\
   \lambda_i &=& \alpha\beta\nu\,\omega_i
 \end{eqnarray}
Furthermore, the boundary $\partial{\cal P}$ is given by $\beta=1$. In this
case, the unit vector $\vec \xi$ represents the touching point (or one of
the touching points in more degenerate cases) between the unit sphere and
its image. Let $\Phi\in \partial{\cal P}$, so $\beta=1$. Then it is easy to
check that
$w_0=\alpha\nu^2$
makes $Q$ positive semidefinite since it
permits a Cholesky decomposition $Q=RR^T$ into a triangular
matrix with a zero on the diagonal:
\begin{equation}
  \label{eq:16}
  R=
  \begin{pmatrix}
    0&0&0&0\\
  -\omega_1\xi_1 \mu_1& 
 \nu \mu_1 & 0&0\\
  -\omega_2\xi_2 \mu_2& 
 0& \nu \mu_2 & 0\\
  -\omega_3\xi_3 \mu_3& 
0&0& \nu \mu_3 \\
  \end{pmatrix} 
\end{equation}
where $\mu_i = \sqrt{\alpha(1-\alpha  w_i^2)}$.
Furthermore, ${\bf n}=
(1,\frac{1}{\nu}\xi_i \omega_i  )$ is a lightlike vector in $\Ker Q$. 

 In the general case $\beta<1$ we have
  \begin{equation}
    \label{eq:12}
    Q^{\Phi_\beta}_w  = \beta^2\; Q^{\Phi_{(\beta=1)}}_{w\beta^{-2}}
+(1-\beta^2)
\begin{pmatrix}
  1&0&0&0\\
  0&0&0&0\\
  0&0&0&0\\
  0&0&0&0\\
\end{pmatrix}
  \end{equation}
Therefore, $Q^{\Phi_\beta}_{w_0 \beta^2}$ as a sum of two positive
semidefinite terms is either positive semidefinite or positive definite.
In the first case we are done with $w_0=\alpha\beta^2\nu^2$. In the other
case we must adjust $w_0$. 
 It is clear that $Q$ is
not positive for $w\rightarrow \pm \infty$. Therefore due
 to continuity,  we can make $Q$ positive semidefinite and degenerate
by increasing or decreasing $w$. Let $w_1<w_2$  
be the points of degeneration  
 and ${\bf n}_1, {\bf n}_2$ corresponding vectors in $\Ker Q_{w_{i}}$. Then
(eq.~(\ref{eq:15})) ${\bf n}_1 Q_0 {\bf n}_1 = w_1 {\bf n}_1^2$ and 
${\bf n}_2 Q_0 {\bf n}_2 = w_2 {\bf n}_2^2$. Furthermore, no nonzero vector 
can be both in $\Ker Q_{w_1}$ and $\Ker Q_{w_2}$. So,
${\bf n}_1 Q_0 {\bf n}_1 > 
  w_2 {\bf n}_1^2$ and $ {\bf n}_2 Q_0 {\bf n}_2 > 
  w_1 {\bf n}_2^2$, providing $(w_2-w_1) {\bf n}_1^2 <0$ and 
 $(w_2-w_1) {\bf n}_2^2 >0$.
Therefore, increasing $w$ will make $\Ker Q$
time-like and decreasing $w$ will make it space-like.
This proofs the claim of Theorem~\ref{thm2}, existence of a suitable $w_0$.
Uniqueness can be shown easily. It also  follows indirectly 
from the uniqueness of the convex roof extension, 
Theorem~\ref{thmRoof}.
More details can be found in \cite{hellmund08}.

\section{Explicit examples}
Let us demonstrate our construction   on some examples.

\subsection{Bistochastic maps or unital  channels}

Unital 1-qubit channels are quite trivial. We have $\vec t=0, {\bf\Lambda}
=\diag(\lambda_1,\lambda_2,\lambda_3)$  and therefore
$w=\max(\lambda_1^2,\lambda_2^2, \lambda_3^2)$  fulfills the conditions of
Theorem~\ref{thm1} and provides the  roof

\begin{equation}
  \label{eq:9}
 C(\rho)= q^{1/2}(\rho) =\sqrt{ (1-w)x_0^2 +
\sum_{i=1}^3 \left( w-\lambda_i^2\right) x_i^2}
\end{equation}
which is flat in one direction since one of the terms in the sum vanishes.

Nevertheless,  this case includes channels of all Kraus lengths
between 1 and 4.

\subsection{Axial symmetric channels}
Every channel commuting with  rotations about 
the $x_3$-axis is of the form 
\begin{equation}
  \label{eq:11}
  \Phi(\rho) =
\begin{pmatrix}
\alpha x_{00}  +(1-\gamma) x_{11}     & \beta x_{01} \\
\beta x_{10} &   \gamma x_{11} + (1 - \alpha) x_{00}
\end{pmatrix}.
\end{equation}
This corresponds to
${\bf\Lambda}=\diag(\beta,\beta,\alpha+\gamma-1)$ and $\vec t=
(0,0,\alpha-\gamma)$. 
This family includes many standard channels,
e.g.,
\begin{itemize}
 \item the amplitude-damping channel (length 2, non-unital) for $\gamma=1,\;
\beta^2=\alpha$;
\item the phase-damping channel (length 2, unital) for
$\alpha=\gamma=1$ and
\item the depolarizing channel (length 4, unital) for $\alpha=\gamma,
\; \beta=2\alpha-1$.
\end{itemize}
Here we find that $q_w^{1/2}$ is a convex roof for
$ w = \max(\beta^2,\beta_c^2)$ with
\begin{equation}
  \label{eq:14}
  \beta_c^2 = 1+2\alpha\gamma-\alpha-\gamma-
2\sqrt{\alpha(1-\alpha)\gamma(1-\gamma)}.
\end{equation}
In the case $\beta^2\ge \beta_c^2$ we have $\Ker Q =\text{Span}\{e_x,e_y\}$
and the resulting roof is flat. In the other case we have
a one-dimensional $\Ker Q$ generated by ${\bf n}=(1,0,0,z_0)$ with
$z_0=\frac{\sqrt{\gamma(1-\gamma)}+\sqrt{\alpha(1-\alpha)}}
{\sqrt{\gamma(1-\gamma)}-\sqrt{\alpha(1-\alpha)}}$ and a non-flat roof.

\section{Conclusion}

We calculated the concurrence $C_\Phi$ of all trace-preserving positive
1-qubit maps and therefore for general $2\times n$ bipartite systems 
with rank-2 input states. 

The concurrence is
real linear on each member of a unique bundle of straight
lines crossing the Bloch ball. The bundle consists either
of parallel lines or the lines meet at a pure state, or
they meet at a point  outside the Bloch ball. Furthermore,
$C_{\Phi}$ turns out to be the restriction of a Hilbert
semi-norm to the state space.

More details and applications, including the entanglement of formation in 
$2\times n$ systems 
and the Holevo capacity \cite{schumacher97}
of channels will be given in \cite{hellmund08}.


\end{document}